# Optical quenching and recovery of photoconductivity in single-crystal diamond

J. Chen[1,a)], S. Lourette[2], K. Rezai[2], T. Hoelzer[3], M. Lake[4], M. Nesladek[5], L.-S. Bouchard[4], P. Hemmer[1], and D. Budker[2,6]

[1]*Department of Electrical and Computer Engineering, Texas A&M University, College Station, TX 77843, USA*
[2]*Department of Physics, University of California, Berkeley, CA 94720, USA*
[3]*Department of Physics, Rheinisch-Westfälische Technische Hochschule (RWTH) Aachen University, 52062 Aachen, Germany*
[4]*Department of Chemistry and Biochemistry, University of California, Los Angeles, CA 90095-1569, USA*
[5]*Institute for Materials Research (IMO), Hasselt University, Belgium*
[6]*Helmholtz Institute, Johannes Gutenberg University, 55099 Mainz, Germany*

We study the photocurrent induced by pulsed-light illumination (pulse duration is several nanoseconds) of single-crystal diamond containing nitrogen impurities. Application of additional continuous-wave light of the same wavelength quenches pulsed photocurrent. Characterization of the optically quenched photocurrent and its recovery is important for the development of diamond based electronics and sensing.

With unique properties such as a wide bandgap, high thermal conductance and broadband optical transmittance, diamond has found a broad range of applications in opto-electronics including electron emitters [1], windows for high-power lasers [2], and x-ray detectors [3]. Diamond is an excellent photoconductor and it can be used for ultrasensitive UV detectors [4]. Photocurrent can also be induced by photon absorption on impurities and defects [5,6,7], which promotes electrons to the conduction band. In general, the diamond's photoelectric properties, such as optical absorption spectra [8,9] and photon-to-electron quantum efficiency, depend strongly on the doping impurities [10,11]. In addition, diamond impurities like nitrogen vacancy (NV$^-$) centers have applications in the research for quantum computing [12-14], quantum optics [15], quantum electronics [16,17], as well as in sensing [18]. Measuring photocurrent will aid in our understanding of the defects' energy levels and provide an insight into diamond's photoelectric properties.

Optical quenching of photocurrent in diamond was first discussed in boron doped epitaxial diamond films grown on nitrogen-rich type Ib diamond with continuous wave (CW) deep ultraviolet (DUV) excitation light and white quenching light [18]. The CW DUV light and white light with non-overlapping spectra play distinctive roles with regards to mobile electrons and holes, and thus result in excitation or quenching of photocurrent. However, there is no previous report of excitation-quenching behavior of photocurrent at the same wavelength in diamond. Recently, photocurrent detection of magnetic resonance (PDMR) in the ground-state of the NV$^-$ centers was reported [19], however, with a large background unrelated to the NV$^-$ centers. The investigation of whether the optical quenching effect reported here may be used to increase the contrast of PDMR, as well as the search for NV$^-$ centers' metastable state level, open the way to practical quantum applications of this technique.

Here, we present our investigation of optical quenching and recovery of photocurrent in bulk diamonds. Continuous-wave 532 nm laser light was found to quench the photocurrent produced by pulsed lasers of various wavelengths between 532 nm (inclusive) and 660 nm. In addition, the residual (incompletely quenched) photocurrent was found to decrease with increasing CW laser intensity until saturation. When the CW laser was removed, the residual photocurrent gradually recovered to the unquenched level with a recovery time that was found to depend on bias voltage, pulsed-laser wavelength, and pulsed-laser intensity.

_________________________

a)Electronic mail:   tideecho@gmail.com

In this report, single-crystal bulk diamonds of different types were studied: chemical vapor deposition (CVD) IIa, and high temperature high pressure (HTHP) Ib with different concentrations of nitrogen-vacancy centers (NV$^-$). The nitrogen and NV$^-$ concentrations of the four samples studied are presented in Table 1. The primary sample studied is a CVD IIa single-crystal bulk diamond with intrinsic nitrogen concentration of < 1 ppm (Sample 1). Samples 1, 2 and 4 were irradiated with relativistic electrons and annealed with the initial goal of creating NV$^-$ centers; we note that such treatment reduces the concentration of interstitial defects [20] and suppresses surface conduction [21].

Table.1 Samples of single-crystal diamond used in this work.

Nitrogen concentration, NV concentration, integrated photocurrent, quenching ratio. We define quenching ratio to be the maximal observed reduction in photocurrent divided by unquenched photocurrent. The values for integrated photocurrent per pulse and quenching ratio were measured with the 532 nm pulsed laser at 20 µJ per pulse, with a DC bias voltage of 80 V. Samples 3 and 4 are from the same batch of diamonds, but sample 4 was irradiated with electrons and annealed to produce NV centers.

|  | [N] in ppm | [NV$^-$] in ppb | Integrated photocurrent per pulse (pC) | Quenching ratio |
| --- | --- | --- | --- | --- |
| Sample 1 | < 1 | 3 | 2 | 87% |
| Sample 2 | < 1 | 11 | 5 | 90% |
| Sample 3 | 100 | < 1 | 7 | 72% |
| Sample 4 | 100 | 5000 | 0.5 | < 20% |

For each sample, a pair of titanium/gold electrodes was deposited onto the surface with a 20 µm gap and a bias voltage of up to 80 V was applied across the electrodes. The photocurrent was either excited with a 532 nm pump pulsed laser (Spectra Physics Q-switched Nd-YAG laser DCR-11 with ~10 ns pulses at 10 Hz repetition rate) or with a tunable dye laser (Quanta Ray PDL-2; pumped with the Nd-YAG laser). The pulsed-laser beams were focused to a plane slightly above the surface of the diamond, such that the beam nearly fills the 20 µm gap between the electrodes. In addition, a collimated CW laser with a beam diameter of 0.5 mm was applied to the entire region as shown in Fig. 1(a). When appropriate, a fast current pre-amplifier (Stanford Research Systems SR445) was used to enhance the signal-to-noise ratio of the detected photocurrent. The signal is then recorded on an oscilloscope.

Illumination with 532 nm 10 nanosec pulsed light produces an electric response of ~30 ns duration; the relationship between the integrated current and pulse energy is shown in Fig. 1(b). When the sample was co-illuminated with both an unfocused 532 nm CW laser light at 1.3 W/mm$^2$ and a 532 nm pulsed laser at 0.14 mJ per pulse, the detected photocurrent was reduced (quenched) by up to an order of magnitude, as shown in Fig. 1(c). This order-of-magnitude reduction in photocurrent is only observed in three of the four diamonds studied (see Table 1), the exception being the type Ib diamond with many NV centers (sample 4). Because samples 3 and 4 are from the same diamond batch, the difference between samples 3 and 4 in photocurrent and quenching can be primarily attributed to irradiation with electrons and annealing, which were only performed on sample 4. During the irradiation lattice defects are produced, of which part is removed by subsequent annealing that recombines vacancies with N to NV centers [22]. We also note that that HTHP diamonds (samples 3 and 4) require significantly less precise positioning of the pulsed beam between the electrodes in order to produce photocurrent than what is required for the CVD diamonds.

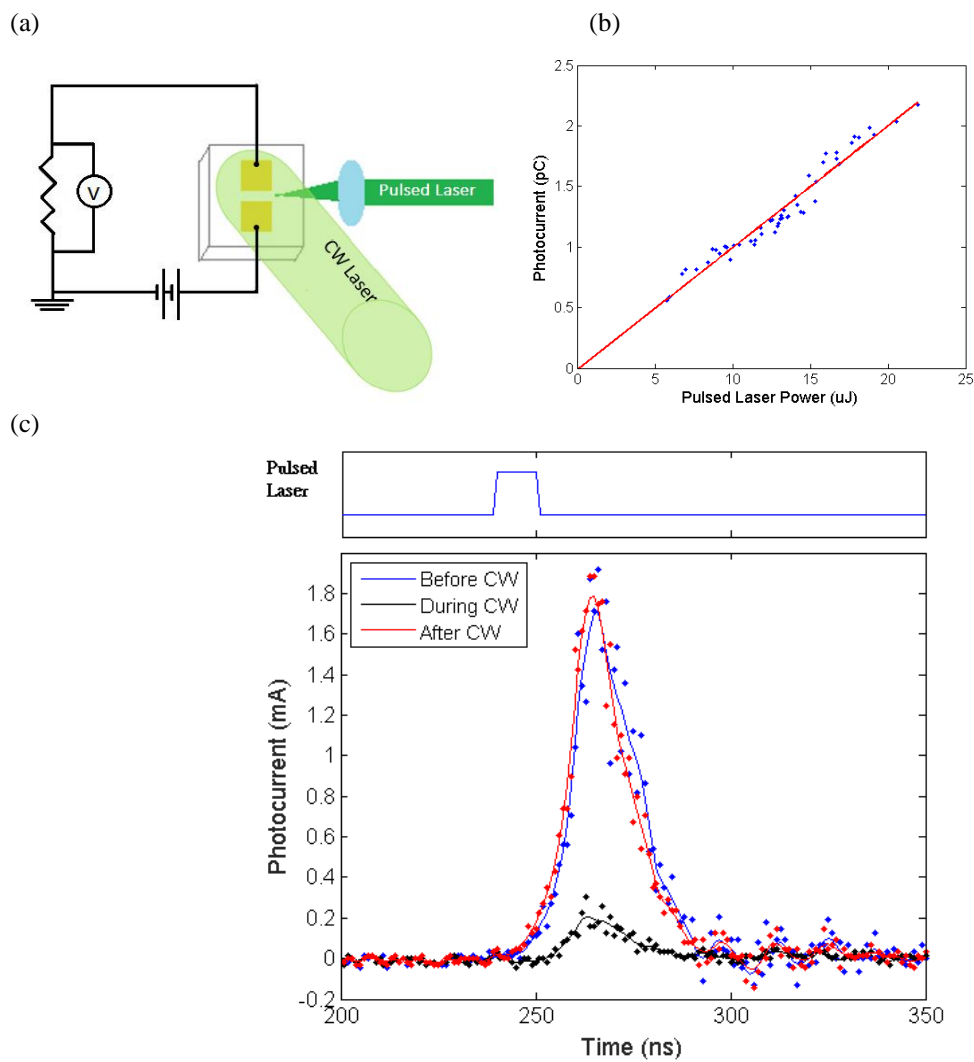

Fig. 1. (a) Experimental setup. The dye-laser light or the light from the doubled YAG was focused through a lens to the side of the diamond where the electrodes were deposited so as not to illuminate the electrodes, while a collimated 180 mW CW 532 nm laser light with a beam diameter of 0.5 mm directly illuminated the entire electrode region. (b) Photocurrent amplitude vs. 532 nm pulsed laser intensity of sample 1. (c) Time trace of pulsed photocurrent in sample 1 produced by illumination with a 0.14 mJ 532 nm pulsed laser. Blue line: photocurrent with 532 nm pulsed light before CW 532 nm light illumination. Black line: photocurrent with both 532 nm pulsed light and CW 532 nm co-illumination. Red line: photocurrent with 532 nm pulsed light after removing CW 532 nm light illumination.

To investigate the dynamics of optical quenching of photocurrent, we use a quenching/recovery sequence, whose timing diagram is shown in Fig. 2 (top). Initially, the photocurrent produced by the 532 nm pulsed laser (140 µJ) alone is monitored. Then after 10 s the CW laser is switched on, and after another 20 s it is switched off. During the entire sequence, a bias voltage of 60 V is applied. The oscilloscope trace is recorded each second (10 pulses), and integrated to obtain an average integrated photocurrent. We perform this procedure on sample 1 with four different CW laser powers to obtain the plot shown in Fig. 2 (bottom). The electrical background signal produced by the Q-switch trigger pulse is subtracted from the signals. As shown in the plot, the photocurrent is quenched during CW laser illumination and gradually recovers after the CW laser is removed. The residual photocurrent decreases with increasing CW laser power over the range of powers studied, as shown in Fig. 2 (Inset).

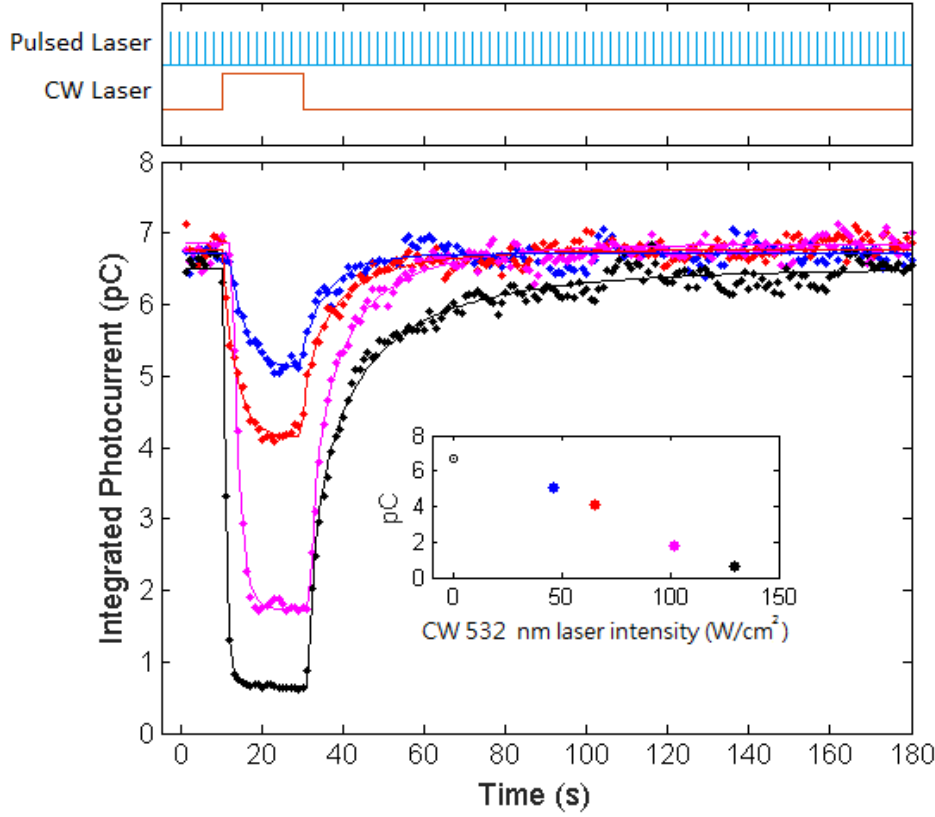

Fig. 2. Optical quenching of pulsed photocurrent with CW light. (Top): Timing for the photocurrent-suppression and recovery measurement. The 532 nm pulsed laser remained on throughout this experiment, whereas the 532 nm CW laser was only on for 20 s. (Bottom): Time-integrated pulse photocurrent (in charge units) versus elapsed time for different CW light powers intensities in sample 1: Blue = 46 W/cm2, Red = 65 W/cm2, Magenta = 102 W/cm2, Black = 130 W/cm2. Photocurrent suppression by an order of magnitude is observed at the highest laser power. (Inset): Dependence of quenched pulse photocurrent on CW light power.

In the case of the photocurrent in a sample with low absorption, the photocurrent density produced through 1 photon ionization may be expressed as [23]

$$j = e\mu\tau\eta\varphi(1-R)\alpha V/d \qquad (1)$$

, where $e$ the unit charge, $\mu$ is the carrier mobility, $\tau$ is the excitation life time, $\eta$ is the photon-to-electron quantum efficiency, $\varphi$ is photon flux, $R$ is reflectivity, $\alpha$ is the absorption coefficient, $V$ is the bias voltage, and $d$ is the electrode gap separation. The photocurrent in our samples is found to be linear in bias voltage, and we obtain very rough estimates for quantum efficiency of 0.5 for sample 1 and 0.2 for sample 2 using our experimental parameters and literature values for absorption of 2 cm$^{-1}$ [24] and for the mobility-lifetime product of $10^{-4}$ cm$^2$/V for type IIa diamond [25].

The relationship that we observe between photocurrent and CW laser power might be explained by a physical model: electrons are photo-excited from electron donor defects, such as substitutional nitrogen (P1 centers), to the conduction band excited by photons provided by the CW laser, then fall into trap defects such as neutral nitrogen-vacancy centers (NV$^0$) in addition to ionized donor defects (N$^+$). The pulsed laser does likewise, but also excites electrons from the traps to the conduction band via a short-lived intermediate state (in the case of NV$^-$, this would be the excited triplet state). In the case of NV triplet state 2 photon ionization is needed requiring high laser power [19]. The CW laser lacks the intensity to excite significant numbers of electrons from the trap state at the same rate as the pulse laser, so between the pulses electrons are shifted from donor states to trap states, thus producing the quenching effect that we observe. With our experimental settings, the intensity of the light from the pulse laser for the duration of the pulse is 7-8 orders of magnitude higher than the light intensity from the CW laser, though the total charge moved per second is comparable. The optical quenching of the photocurrent increases with increasing CW laser power until saturation is reached (more than 90% quenching based on our data).

The photocurrent recovery process as a function of time after removing the CW 532 nm light was fitted of the form (Kohlrausch function)

$$I(t) = I_0 \left( 1 - e^{-\left(\frac{t}{\tau}\right)^\beta} \right) \qquad (2)$$

, where $I(t)$ is the time dependent photocurrent, $t$ is the elapsed time since the moment when CW light was turned off, $\tau$ is the recovery time, and $\beta$ is a constant. The stretched exponential is consistent with the previous studies of transient photocurrent discharging and charging in diamond [26], as well as fluorescence decay commonly found in some crystalline solids like porous silicon or CdSe-ZnSe [27]. Whether it is due to a time dependent decay rate or a superposition of several exponential decays is still unclear [28]. The fact that the stretched exponential form exists in many materials may suggest a common property, related to charge-trapping kinetics.

The recovery rate $1/\tau$ is plotted as a function of pulsed laser energy [Fig. 3(a)] for 532 nm light. The recovery rate increases with increasing pulse energy, which is consistent with the assumption that the pulsed light depopulates the trap. The photocurrent recovery rate also depends on the DC bias voltage as shown in Fig. 3(c) for two pulsed-laser wavelengths/pulse energies: 532 nm at 70 μJ and 585 nm at 60 μJ. The recovery times at 60 V were 2.4 s and 15 s, respectively, which is a difference much larger than what one would expect from changing only the pulse energy. At 630 nm, the recovery takes several hours. The recovery dynamics were found to obey the stretched-exponential function with $\beta = 0.6\sim1.1$. In general, longer wavelength, smaller pulsed laser energy and voltage results in higher β [Fig. 3 (b)(d)]. We found that the photocurrent recovery time shortens with increasing bias voltage, suggesting that the higher photocurrent at higher bias voltage depopulates the electron traps faster. This is in contrast to previously reported diamond transient photocurrent decay [10,26] that is voltage independent. This discrepancy can be explained if more charge is being trapped at lower electric fields because the travel time between electrodes becomes significant when compared to the recombination lifetime. While the observed recovery time constant was found to strongly depend on both the applied voltage and pulse energy, we did not observe a significant dependence of the recovery time on the CW laser power. As for the quenching-onset dynamics when the CW light is turned on, an exponential fitting of the photocurrent dynamics reveals no significant dependence on either the bias voltage or pulsed laser power; however, the optical quenching-onset time is reduced with increased CW laser power.

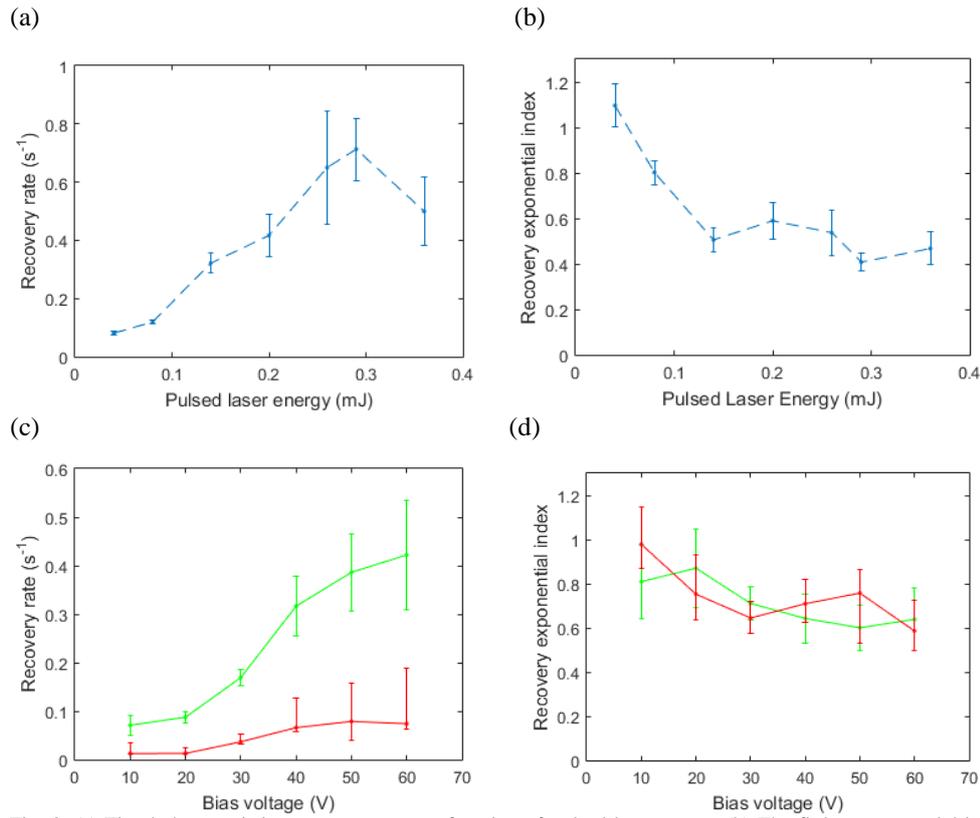

Fig. 3. (a) Fitted characteristic recovery rate as a function of pulsed laser power. (b) The fitting exponential index for pulsed laser energy dependent recovery rate. (c) The characteristic recovery rate vs. bias voltage for 585 nm and 532 nm pulsed lasers of sample 1. Red: 585 nm laser, and green: 532 nm laser. (d) The fitting exponential index for voltage dependent recovery rate.

When a red laser pulse is applied a few nanoseconds after the end of each green laser pulse (532 nm at 70 µJ and 630 nm at 60 µJ ) at a bias of 60 V, the recovery rate is 3.7 times faster than with green pulse excitation alone. Furthermore, applying a red pulse alone gives a slow recovery time of 90 minutes. This can be explained by the proposed model if the trap can only be excited to an intermediate state by green light, but from there it can be excited to the conduction band by either red or green light, as shown in Fig. 4.

In our conduction-band-trap model, we assume that P1 centers are the electron donors because these are the dominant defects in the diamond investigated here. The neutral NV center remains a possible candidate for the dominant electron trap state.

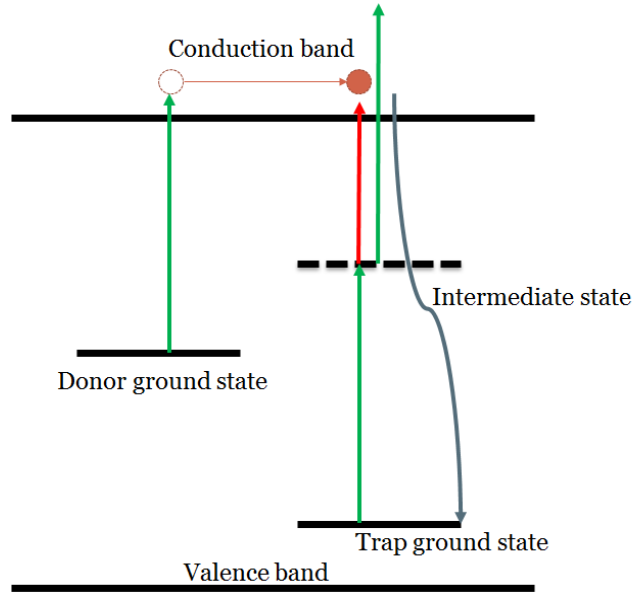

Fig. 4. Proposed model for optical quenching of photocurrent in diamond.

From observation of the four different samples, it is found that photocurrent and quenching can be observed in both type Ib and type IIa diamond, suggesting that they share a common underlying mechanism. Also, we can infer from the large observed differences between samples 3 and 4 that the process through which $NV^-$ centers are produced in the type Ib diamonds greatly reduces both photocurrent and quenching.

To address the possibility that the quenching of photocurrent might actually be due to heating induced by the CW laser, we performed the following experiment: the CW laser and pulsed laser are switched on for 20 s, during which time the photocurrent rapidly decreases to the "quenched" level. Then the two lasers are switched off for a varied delay time, after which the pulsed laser is switched back on and the photocurrent is measured. We observed no change to this photocurrent level at different delay times, which would be expected if the quenching effect were caused by heating. Furthermore, an estimate of the heating caused by the CW laser predicts a slow rate of 0.1 °C per second, which is orders of magnitude smaller than what would be required to form a plausible model where the quenching dynamics are governed by heating effects. Other possible causes of the photocurrent quenching include the shortening of the excited-state lifetime in the presence of CW light analogous to what stronger irradiance does to metal enhanced fluorescence [29], and light-induced change of carrier mobility due to formation of polar traps [30]. However, these possible explanations require further modification and introduction of new assumptions to fully explain all the observed phenomena reported here, specifically the lengthy recovery time.

In summary, we studied optical quenching and recovery of photocurrent in bulk single-crystal diamond. Notably, in the presence of CW light of the same wavelength we observed an order of magnitude reduction of photocurrent. The dependences of the recovery and quenching times on external bias voltage and light intensity were also investigated in this work. Based on these data, we suggest a model in which the electron donor center is ionized by a single green photon, and the electron trap state can in turn be ionized by two photons. Furthermore, the observations are consistent with the trap having an excited state that can be populated by green light, which can then be ionized with red light. For single-crystal diamond, the study of optical quenching of photocurrent paves the way to better understand electronic levels of defects in diamonds. In addition, the investigation of photocurrent dynamic time constants in this report constitutes a new general method to probe photoconductive semiconductors' electric properties.

This work was supported by AFOSR and the DARPA QuASAR program, by NSF Grant No. ECCS-1202258, and by DFG through the DIP program (FO 703/2-1). The authors thank Andrey Jarmola and Pauli Kehayias for their help throughout the project, Professor Irfan Siddiqi for help with the wirebonding of the electrodes, and Professor Marcis Auzinsh for insightful discussions.

[1] T. Yamada, A. Sawabe, S. Koizumi, J. Itoh and K. Okano, Appl. Phys. Lett. 76, 1297 (2000)

[2] D. H. Douglas-Hamilton, E. D. Hoag, and J. R. M. Seitz, J. Opt. Soc. Am. 64(1), 36-38 (1974)

[3] J. Bohon, E. Muller, and J. Smedley, J. Synchrotron Radiat. 17, 711-718 (2010)

[4] J.-F. Hochedez, W. Schmutz, Y. Stockman, U. Schühle, A. BenMoussa, S. Koller, K. Haenen, D. Berghmans, J.-M. Defise, J.-P. Halain, A. Theissen, V. Delouille, V. Slemzin, D. Gillotay, D. Fussen, M. Dominique, F. Vanhellemont, D. McMullin, M. Kretzschmar, A. Mitrofanov, B. Nicula, et al, Adv. Space Res. 37(2), 303-312 (2006)

[5] J Rosaa, M Vaněčeka, M Nesládekb, L.M Stalsb, Diam. Relat. Mater. 8(2), 721-724 (1999)

[6] M. Nesládek, K. Meykens, K. Haenen, L. M. Stals, T. Teraji, and S. Koizumi, Phys. Rev. B 59, 14852 (1999)

[7] E. Rohrer, C.E. Nebel, M. Stutzmanna, A. Flöter, R. Zachai, X. Jiang, C.-P. Klages, Diam. Relat. Mater. 7(6), 879-883 (1998)

[8] M. Nesládek, K. Meykens, L. M. Stals, M. Vaněček, and J. Rosa, Phys. Rev. B 54, 5552 (1996)

[9] Z. Remes, R. Petersen, K. Haenen, M. Nesladek, M. D'Olieslaeger, Diam. Relat. Mater. 14(3-7), 556-560 (2004)

[10] J. A. Elmgren and D. E. Hudson, "Imperfection Photoconductivity in Diamond," Phys. Rev. 128, 1044 (1962)

[11] R. G. Farrer and L. A. Vermeulen, J. Phys. C: Solid State Phys. 5(19), 2762 (1972)

[12] T. H. Taminiau, J. Cramer, T. van der Sar, V. V. Dobrovitski & R. Hanson, Nat. Nanotechnol. 9, 171-176 (2014)

[13] F. Dolde, V. Bergholm, Y. Wang, I. Jakobi, B. Naydenov, S. Pezzagna, J. Meijer, F. Jelezko, P. Neumann, T. Schulte-Herbrüggen, J. Biamonte & J. Wrachtrup, Nat. Commun. 5, 3371 (2014)

[14] J. Scheuer, X. Kong, R. S. Said, J. Chen, A. Kurz, L. Marseglia, J. Du, P. R. Hemmer, S. Montangero, T. Calarco, B. Naydenov and F. Jelezko, New J. Phys. 16, 093022 (2014)

[15] H. Bernien, L. Childress, L. Robledo, M. Markham, D. Twitchen, and R. Hanson, Phys. Rev. Lett. 108, 043604 (2012)

[16] N. Mizuochi, T. Makino, H. Kato, D. Takeuchi, M. Ogura, H. Okushi, M. Nothaft, P. Neumann, A. Gali, F. Jelezko, J. Wrachtrup & S. Yamasaki, Nature Photon. 6, 299-303 (2012)

[17] A. Brenneis, L. Gaudreau, M. Seifert, H. Karl, M. S. Brandt, H. Huebl, J. A. Garrido, F. H. L. Koppens & A. W. Holleitner, Nat. Nanotechnol. 10, 135-139 (2015)

[18] M. Liao, Y. Koide, J. Alvarez, M. Imura, and J. Kleider, Phys. Rev. B 78, 045112 (2008)

[19] E. Bourgeois, A. Jarmola, P. Siyushev, M. Gulka, J. Hruby, F. Jelezko, D. Budker, and M. Nesladek, Nat. Commun. 6, 8577 (2015)

[20] D. J. Twitchen, M. E. Newton, J. M. Baker, T. R. Anthony and W. F. Banholzer, J. Phys. Condens. Mat. 13, 2045 (2001)

[21] M. I. Landstrass and K. V. Ravi, Appl. Phys. Lett. 55, 975 (1989)

[22] V. M. Acosta, E. Bauch, M. P. Ledbetter, C. Santori, K.-M. C. Fu, P. E. Barclay, R. G. Beausoleil, H. Linget, J. F. Roch, F. Treussart, S. Chemerisov, W. Gawlik, and D. Budker, Phys. Rev. B 80, 115202 (2009)

[23] C. Nebel, Thin-Film Diamond II: (part of the Semiconductors and Semimetals Series) (Academic Press, 2004)

[24] J. Wilks and E. Wilks, Properties and Applications of Diamond (Butterworth-Heinemann Ltd., 1991)

[25] F. J. Heremans, G. D. Fuchs, C. F. Wang, R. Hanson and D. D. Awschalom, Appl. Phys. Lett. 94, 152102 (2009)

[26] X. Chen, B. Henderson and K. P. O'Donnell, Appl. Phys. Lett. 60, 2672 (1992)

[27] R. Chen, J. Lumin. 102–103, 510-518 (2003)

[28] E. Rohrer, C.F.O. Graeff, C.E. Nebel, M. Stutzmann, H. Giittler, R. Zachai, Mater. Sci. Eng., B 46(1-3), 115-118 (1997)

[29] J. O. Karolin and C. D. Geddes, J. Fluoresc. 22(6), 1659-1662 (2012)
[30] M. Vala, M. Weiter, O. Zmeškal, S. Nešpůrek, P. Toman, Macromol. Symp. 268(1), 125-128 (2008)